\documentclass[9pt,letterpaper,twoside]{article}
\usepackage{ol2}
\usepackage[draft,implicit=false]{hyperref}
\usepackage{amsmath}

\begin{document}

\twocolumn[ 

\title{Extended object reconstruction from image intensities blurred by unknown aberrations}

\author{Yifeng Shao,$^{1,*}$ Niek Doelman,$^{2,3}$ Silvania F. Pereira$^{1}$ and H. Paul Urbach$^{1}$}

\address{
$^1$Optics Research Group, Delft University of Technology, The Netherlands,
\\
$^2$Netherlands Organization for Applied Scientific Research (TNO), The Netherlands,\\
$^3$Leiden Observatory, Leiden University, The Netherlands,\\
$^*$Corresponding author: y.shao@tudelft.nl
}

\begin{abstract}
We describe and experimentally validate an algorithm to reconstruct an unknown extended object from through-focus measured image intensities blurred by unknown aberrations. It is demonstrated by experiment that the method can recover diffraction-limited image quality. The algorithm is a rigorous, computational optics alternative to the conventional adaptive optics which requires additional high-end optical devices to measure and to correct the wavefront aberrations. The method can be applied to any optical system that fulfills the condition of incoherent imaging.
\end{abstract}]









For advanced optical systems operating at the diffraction-limit, aberrations are the main course of image blurring. Usually, dynamic aberrations are caused by nonuniform media e.g.\ biomedical tissue or by perturbation to uniform media e.g.\ turbulence in the air. In adaptive optics, dynamic aberrations are determined by measuring the wavefront in the pupil of the optical system using e.g.\ a Shack-Hartmann wavefront sensor (SHWS) and are corrected in the light path by a deformable mirror (DM) or a spatial light modulator (SLM). Adaptive optics has been studied extensively in astronomy, where a natural star or an artificial laser guide star is used as a point reference source, and is now progressively applied in biomedical imaging. When imaging the eye, it was proposed to use the fluorescence of retina cells for wavefront sensing \cite{Diaz:1999}. In \cite{Tao:2011}, adaptive optics was implemented in a confocal microscopy set-up and wavefront sensing uses the light emitted by fluorescent beads embedded in the sample. 

Most often in biological imaging, aberrations are not determined by direct wavefront sensing but by indirect wavefront sensorless method. For example, in \cite{Girkin:2003,Ji:2010} the point-spread function (PSF) of two-photon microscopy was improved by optimizing the phase in the pupil by a correcting device. The approach in \cite{Booth:2006,Debarre:2007} corrects the aberrations via the optimization of a image quality metric. The wavefront sensorless method requires many sequential measurements and corrections. 

Computational optics can help to improve the image quality while avoiding additional high-end optical devices. The image quality is improved by reconstructing the object by deconvolution. However, because the aberrations are unknown, the PSF, which is the kernel of the convolution operator, cannot be computed. In \cite{Primot:1990,Beverage:2002}, the aberrations are determined by using a SHWS and the PSF is computed based on the determined aberrations. Another type of deconvolution is referred to as blind deconvolution, which uses the maximum entropy method \cite{Ayers:1988,Lam:2000} or the Richards-Lucy algorithm \cite{Fish:1995,Chen:2013} to optimize the PSF together with the object by setting the measured image intensity as optimization constraint. The performance of the blind deconvolution relies on good initial guess of the aberrations. Recently, a novel algorithm was presented to extend the depth-of-focus (DOF) of an optical system by using a micro-lens array \cite{Levoy:2009,Broxton:2013}. The algorithm relates the location of the object in the DOF to defocus aberration, but by principle, the algorithm cannot retrieve other aberrations.

Here we describe a computational imaging algorithm which can reconstruct an unknown extended object from image intensities blurred by unknown aberrations. Neither a SHWS, nor a correcting device like a DM or an SLM will be used. The algorithm requires at least two image intensities measured in the focal region and the locations of the measurement planes must be known precisely. The numerical apertures of the optical system and the illumination wavelength must be known as well.

The method of the algorithm was first proposed by Paxman et al.\ in \cite{Paxman:1992} and was generalized by Vogel et al.\ in \cite{Vogel:1998}. The described algorithm is similar to the one recently studied in \cite{Liu:2012}, except that regularization is not introduced appropriately. Furthermore, the mentioned paper simulates the image intensity by convoluting the object with the measured intensity of the PSF.  In this paper, we introduce Tikhonov's regularization to deal with the noise and choose the parameter of Tikhonov's regularization using the L-curve method proposed by Hansen \cite{Hansen:1999}. We demonstrate by experiment that diffraction-limited image quality can be recovered by the algorithm for a severely aberrated optical system. 

\begin{figure}[htb]
\fbox{\includegraphics[width=\linewidth]{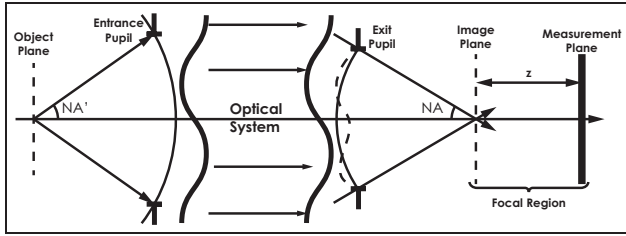}}
\caption{The sketch of the optical system. The optical system is treated as a black-box bounded by the entrance pupil and the exit pupil.}
\label{fig:optical_system}
\end{figure}

The optical system we study is sketched in Figure \ref{fig:optical_system}.  The location of the best nominal image plane is associated to the location of the object plane. $z$ is the distance relative to the best nominal image plane. We denote $\mathbf{s}$ as the spatial frequency coordinate in Fourier space. Suppose that $\mathrm{O}(\mathbf{s})$ is the Fourier transformed intensity in the object plane. For incoherent imaging, the Fourier transformed intensity in the $z$-plane is given by
\begin{equation}\label{eq:intensity_formation}
  \mathrm{I}_z(\mathbf{s}) = \mathrm{O}(\mathbf{s})\mathrm{H}_z(\Phi;\mathbf{s}),
\end{equation}
where $\mathrm{H}_z(\Phi;\mathbf{s})$ is the Fourier transformed intensity of the PSF in the $z$-plane. We compute $\mathrm{H}_z(\Phi;\mathbf{s})$ using the Debye diffraction integral \cite{Wolf:1959} and evaluate it using the chirp z transform \cite{Rabiner:1969} (see supplementary material). \eqref{eq:intensity_formation} tells that in Fourier space, the spatial frequency components of $\mathrm{O}(\mathbf{s})$ will be modulated by $\mathrm{H}_z(\Phi;\mathbf{s})$, which is determined by the aberrations $\Phi$. 

Let $\ell$ be the index of the measurement plane located at $z_\ell$. We define an error functional in Fourier space by
\begin{align}\label{eq:error_function1}
   {\cal  L}_\gamma[\Phi,\mathrm{O}(\mathbf{s})]
  =& \iint\nolimits_{|\mathbf{s}|\leq\frac{2\text{NA}}{\lambda}}
  \sum\nolimits_\ell\left|\mathrm{I}_{\ell}(\mathbf{s})-
  \mathrm{O}(\mathbf{s})\mathrm{H}_{\ell}(\Phi;\mathbf{s})\right|^2
  \mathrm{d}\mathbf{s} \\ \nonumber
  +&\gamma\iint\nolimits_{|\mathbf{s}|\leq\frac{2\text{NA}}{\lambda}}
  \left|\mathrm{O}(\mathbf{s})\right|^2\mathrm{d}\mathbf{s}.
\end{align}where $\gamma$ is the parameter of Tikhonov's regularization. Assuming the aberrations $\Phi$ are known, the Fourier transformed object is chosen such that the error functional $\mathrm{O}(\mathbf{s}) \mapsto {\cal L}_\gamma[\Phi, \mathrm{O}(\mathbf{s})]$ is extremum, which yields:
\begin{equation}\label{eq:object}
  \mathrm{O}_\gamma(\Phi;\mathbf{s}) = \frac{\sum_\ell\mathrm{I}_\ell(\mathbf{s})
  \mathrm{H}_\ell(\Phi;\mathbf{s})^*}
  {\sum_\ell |\mathrm{H}_\ell(\Phi;\mathbf{s})|^2+\gamma},\ \ 
  |\mathbf{s}|\leq\frac{2\text{NA}}{\lambda},
\end{equation}
After the aberrations $\Phi$ are retrieved, the Fourier transformed object can be reconstructed by using \eqref{eq:object} for the provided value of $\gamma$. The algorithm uses the L-curve method \cite{Hansen:1999} to choose the value of $\gamma$  (see supplementary material). By substituting \eqref{eq:object} into the error functional \eqref{eq:error_function1}, we eliminate the Fourier transformed object and we obtain:
\begin{equation}\label{eq:error_function2}
 {\cal  L}_\gamma(\Phi) =   \iint\limits_{|\mathbf{s}|\leq\frac{2\text{NA}}{\lambda}}
 \Bigg[\sum\nolimits_\ell |\mathrm{I}_\ell(\mathbf{s})|^2
 -\frac{|\sum_\ell\mathrm{I}_\ell(\mathbf{s})
 \mathrm{H}_\ell(\Phi;\mathbf{s})^*|^2}{\sum_\ell |\mathrm{H}_\ell(\Phi;\mathbf{s})|^2+\gamma}\Bigg]\mathrm{d}\mathbf{s},
\end{equation}
which is non-linear in the aberrations $\Phi$. To retrieve the aberrations $\Phi$, we expands the aberrations by e.g.\ the Zernike polynomials or the Legendre polynomials and formulate an optimization problem for the expansion coefficients using \eqref{eq:error_function2}. (see the supplementary material.)

\begin{figure}[t]
\centering
\fbox{\includegraphics[width=\linewidth]{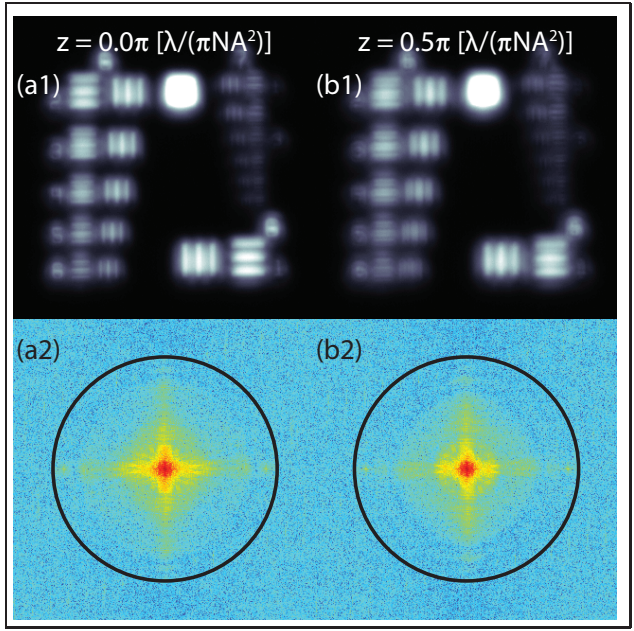}}
\caption{The measured intensities (a1 and b1) and their Fourier transforms (a2 and b2). The cut-off spatial frequency of the optical system is depicted by the circle in the Fourier transforms. Slight but crucial difference between the measured intensities and the Fourier transforms due to different locations of the measurement planes can be observed in the figure.}
\label{fig:result_reference_1}
\end{figure}

\begin{figure}[t]
\centering
\fbox{\includegraphics[width=\linewidth]{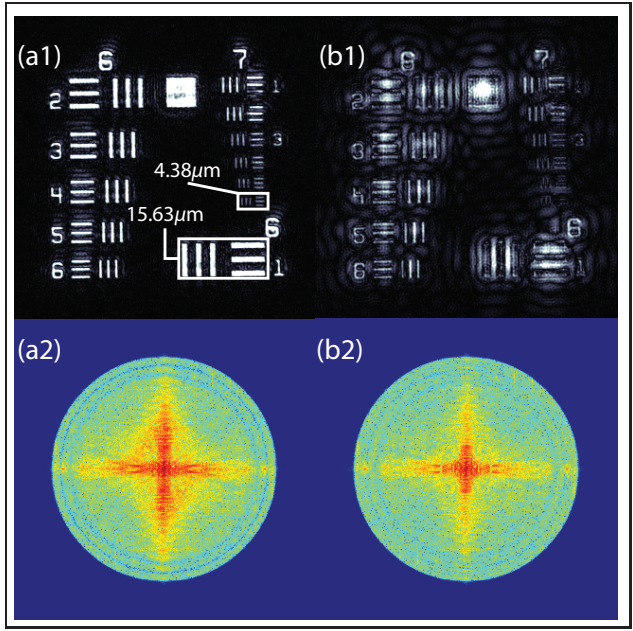}}
\caption{Objects reconstructed using measured intensities (a1 and b1) and Fourier transforms (a2 and b2). On the left: the result is based on the retrieved aberrations. On the right: the result is based on the dominant defocus aberration by deleting other aberrations. The comparison indicates that the improvement of image quality is a consequence of the recovery of the object's spatial frequency components. The recovery relies on the successful retrieval of the aberrations of the microscope objective. The value of $\gamma$ is chosen to be $\gamma = 2.1544$ by the L-curve method. The resolution of the microscope is $0.61\lambda\approx3.18(\mu\text{m})$}
\label{fig:result_reference_2}
\end{figure}

The computational imaging algorithm is validated by a proof-of-principle experiment. The experimental set-up is explained in the supplementary material. The object is a 1951 USAF resolution test target, which is illuminated by monochromatic light at wavelength 625nm. We used a microscope objective (object NA'=0.12 and image NA=0.03) to image the intensity in the object plane to the best nominal image plane. The CCD camera for measuring the intensities has pixel size $4.65\mu\text{m}$ and pixel number $320$-by-$320$ (cropped from total pixel number $1024$-by-$780$). We measure intensities in the best nominal image plane at $z=0.0\pi\ [\lambda/(\pi\text{NA}^2)]$ and in the plane at $z=0.5\pi\ [\lambda /(\pi \text{NA}^2)]$. We introduce defocus aberration to the microscope objective by moving the resolution test target to a defocused position and guarantee that defocus aberration is dominant. The measured intensities will be blurred by the aberrations of the microscope objective as shown in Figure \ref{fig:result_reference_1} (a1), (b1). In Fourier space, the spatial frequency components of the object are modulated by the aberrations and are corrupted by the spatial frequency components of the noise as shown in Figure \ref{fig:result_reference_1} (a2), (b2).

In the experiment, we expand the primary aberrations of the microscope objective (including spherical aberration, coma, astigmatism and defocus aberration) by the fist 15 Zernike polynomials and solve numerically the optimization problem for the expansion coefficients (Zernike coefficients) by a quasi-Newtonian type routine: the "fminunc" function in Matlab (R), which consumes about one minute on a personal computer (Processor: Intel(R) Core(TM) i5-3470 CPU @ 3.20GHZ). The speed is relevant to not only the sampling of the measurement plane, but also the sampling of the exit pupil. We can accelerate the speed by implementing parallel computation. We reconstruct the object based on the retrieved aberrations using \eqref{eq:object}. The $\gamma$ parameter of Tikhonov's regularization is chosen to be $\gamma = 2.1544$ by the L-curve method. Although the L-curve method is an automated method, we can also seek manually in the vicinity of the chosen $\gamma$.

The reconstructed object is shown in Figure \ref{fig:result_reference_2} (a1). We can see that the improvement of image quality is dramatic comparing to the blurred measured intensities shown in Figure \ref{fig:result_reference_1} (a1), (b1). The reconstructed object is sharp and clear, in which we can observe all the elements of the resolution test target: from the largest period $15.63\mu\text{m}$ to the smallest period $4.38\mu\text{m}$. By comparing the Fourier transform of the reconstructed object in Figure \ref{fig:result_reference_2} (c2) with the Fourier transforms of the measured intensities in Figure \ref{fig:result_reference_1} (a2), (b2), we conclude that the algorithm recovers all of the spatial frequency components present in the object. As it is indicated by the comparison in Figure \ref{fig:result_reference_2}, the recovery relies on the successful retrieval of the aberrations of the microscope objective. The value of $\gamma$ also influences significant the reconstruction of the object, which is investigated in the supplementary material. 

\begin{figure}[t]
\centering
\fbox{\includegraphics[width=\linewidth]{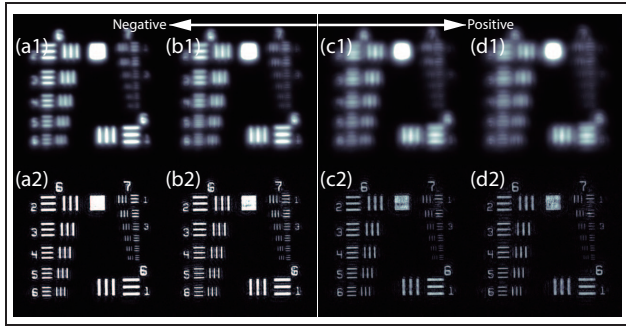}}
\caption{The intensities measured in the best nominal image plane (top row) and the reconstructed objects (bottom row) corresponding to the 4 locations of the resolution test target. Along the negative direction the intensities (a1) and (b1) are less blurred (closer to the object plane) and along the positive direction the intensities (c1) and (d1) are more blurred (further from the object plane).}
\label{fig:result_all}
\end{figure}

\begin{figure}[t]
\centering
\fbox{\includegraphics[width=\linewidth]{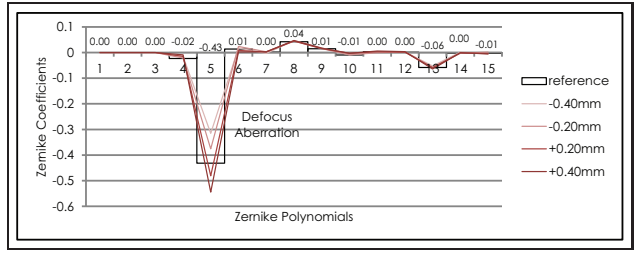}}
\caption{The retrieved aberrations corresponding to the 4 locations of the resolution target (plotted by lines). We plot the retrieved aberrations corresponding to the reference by bars for comparison. As we expected, when moving the resolution test target, defocus aberration changes while other aberrations keeping almost constant.}
\label{fig:retrieval_aberrations}
\end{figure}

To further validate the algorithm, we set the current location of the resolution test target (at a defocused position) as the reference and move the resolution test target in both directions of the optical axis. For 4 different locations of the resolution test target, separated by $0.2\text{mm}$ between the neighbouring locations, we repeatedly measure intensities in the best nominal image plane at $z=0.0\pi[\lambda/(\pi\text{NA}^2]$ and in the plane at $z=0.5\pi[\lambda/(\pi\text{NA}^2]$. We also repeat the procedure of retrieving the aberrations and reconstructing the object based on the retrieved aberrations, for which we use the same value of $\gamma$ as we assume the same signal-to-noise ratio of the measured intensities.

Figure \ref{fig:result_all} (a1-d1) shows that the blurry of the intensities measured in the best nominal image plane changes when we move the resolution test target. We listed the retrieved aberrations corresponding to the 4 locations of the resolution target in Figure \ref{fig:retrieval_aberrations}, which shows that the change of the blurry is in accordance with the change of defocus aberration. In Figure \ref{fig:retrieval_aberrations}, we can observe the existence of not only the dominant defocus aberration, but also other aberrations. The observation matches our expectation that when we move the resolution test target, we change defocus aberrations while keeping other aberrations almost constant. Therefore, we can prove that the algorithm can retrieve the aberrations of the microscope objective. We show the reconstructed objects corresponding to the 4 locations of the resolution test target in Figure \ref{fig:result_all} (a2-d2) and we find that all of them can recover diffraction-limited image quality. We normalize the intensities of the reconstructed objects by the maximum intensity of the one corresponding to the reference location. We can see that the brightness of the reconstructed objects is linked to defocus aberration, but the resolution of the reconstructed objects is not affected by the brightness. 


We discuss the performance of the algorithm from the perspective of spatial frequency components in Fourier space. As shown in Figure  \ref{fig:result_reference_1} (a2), (b2), the Fourier transforms of the measured intensities can be regarded as summations of the spatial frequency components of the object (modulated by the aberrations) and the spatial frequency components of the noise. We can observe that the object components corrupted by the noise components form the blue background, whereas the remaining object components form the red-yellow-green pattern.

For the retrieval of the aberrations, we remark that for a particular object, the modulation by the aberrations may not be revealed by its Fourier transform. An example of such particular object is the periodic object whose Fourier transform consists of scattered spatial frequency components. Therefore, we claim that the algorithm retrieves not the aberrations existing in the optical system, but the aberrations blurring the measured intensities. The signal-to-noise ratio of the measured intensities also plays a role, because the corrupted object components cannot contribute to the retrieval of the aberrations. The algorithm actually utilizes the difference between the remaining object components of the measured intensities.

When reconstructing the object by using \eqref{eq:object}, the corrupted object components will induce artifacts in Fourier space. So, we introduce Tikhonov's regularization to the algorithm to cancel the induced artifacts. In the supplementary material, we show that when the value of $\gamma$ is chosen properly, we can cancel completely the artifacts and recover the remaining object components. The resolution of the reconstructed object will be limited by the cut-off spatial frequency of the optical system, which is determined by the numerical apertures and the illumination wavelength, and by the signal-to-noise ratio of the measured intensities: the remaining object components can be recovered by the algorithm. However, the corrupted object components will be lost eventually, although the value of $\gamma$ is chosen properly. 

To conclude, in this article, we described a rigorous computational optics algorithm which can be applied to any optical system that fulfills the condition of incoherent imaging. The algorithm requires modest modification to the current set-up of the optical system: we can measure intensities either consecutively by a single CCD camera mounted on a translation stage, or simultaneously by multiple CCD cameras mounted by splitting the light path. We demonstrate by experiment that diffraction-limited image quality can be recovered by the algorithm. This paper paves the way for future applications of computational optics in for example astronomical imaging or biomedical imaging, where the diffraction-limited image quality is demanding. 

\section*{Funding Information}
This research was funded by the Technology Foundation STW (12797) of the Netherlands.

\section*{Acknowledgments}
 
The authors thank Prof.\ Joseph J. M.\ Braat and Lei Wei for fruitful discussions and Rob Pols and Tim Stegwee for their supports in the experiment.\\ 
\\
See Supplement 1 for supporting content

\end{document}